\documentclass[12pt,twoside,a4paper,amsmath,amssymb,apsmy,showkeys,showpacs,nofootinbib]{revtex4}

\usepackage[cp1251]{inputenc}
\usepackage[russian,english]{babel}
\usepackage[left=2.5cm, right=2.5cm, top=2.5cm, bottom=2.2cm, bindingoffset=0cm]{geometry}

\usepackage[T2A]{fontenc}

\usepackage{subfig}
\usepackage{graphicx}
\usepackage{graphics}
\usepackage{epsfig}
\usepackage{float}
\usepackage{multirow}
\usepackage{bigstrut}
\usepackage{latexsym}
\usepackage{array}
\usepackage{dcolumn}
\usepackage{bm}

\begin{document}

\thispagestyle{myheadings}

\title{Scattering of relativistic electron beams by the anode mesh in high-current vircators}

\author{Anishchenko S.V.}

\email{sanishchenko@mail.ru}
\affiliation{Research Institute for Nuclear Problems of Belarusian State University \\Bobruiskaya Str. 11, 220030 Minsk, Belarus}

\author{Baryshevsky V.G.}
\email{bar@inp.bsu.by}
\affiliation{Research Institute for Nuclear Problems of Belarusian State University \\Bobruiskaya Str. 11, 220030 Minsk, Belarus}

\author{Gurinovich A.A.}
\email{gur@inp.bsu.by}
\affiliation{Research Institute for Nuclear Problems of Belarusian State University \\Bobruiskaya Str. 11, 220030 Minsk, Belarus}

\begin{abstract}
In a virtual cathode oscillator, the scattering of a high-current relativistic electron beam by the anode mesh leads to formation of an electron cloud near the anode. The cloud particles possess low energy and large spread in velocities caused by multiple scattering and ionization losses. The electrons captured by the cloud do not participate in the oscillations of the virtual cathode and partially block a vircator. As a result, the amplitude of the electric field oscillations is reduced. In order to increase the oscillation amplitude, the thickness of the anode mesh should be equal to the mean free path for electrons in the mesh material.
\end{abstract}


\maketitle

\section{Introduction}
The interaction of charged particles with matter plays an important role in various fields of science and technology. 
Radiation therapy, design of modern particle detectors, and radiation protection of spacecrafts are impossible without a thorough analysis of the passage of charged particles through matter and  quantitative treatment of multiple scattering and energy losses.

The interaction of charged particles with matter is also important in high-current electronics dealing with the propagation of high-power ion and electron beams in electrodynamic structures~\cite{Creedon1990}. Indeed, some of the beam particles, when they inevitably hit one of the structural elements (anode mesh, collector, drift tube, etc.), are reflected due to multiple scattering, return back into the interaction area and continue to interact with electromagnetic fields inside the system. For example, about a quarter of particles normally incident on a steel element of the structure are reflected off. If the incidence angle  significantly differs from normal then the fraction of reflected particles is much higher. The spectrum of kinetic energy for  reflected particles extends from zero to the initial energy. Since the number of incident electrons is comparable with the number of reflected ones, the latter can significantly affect the operation of high-current devices.

A large number of theoretical results devoted to multiple scattering of particles are unsuitable for studying the interaction of high-current electron beams with matter. First, many studies use a small-angle approximation~\cite{Williams1939,GS1940,Moliere1948,Bethe1953,Nigam1959,Marion1967,Voinov1969}. Secondly, the approximate values of elastic electron-atom scattering cross sections differ significantly from those obtained by numerical calculation at energies < 1 MeV~\cite{Mayol1997}. Thirdly, multiple scattering cannot be considered separately from ionization~\cite{Bohr1913,Bethe1932,Bloch1932,BetheAshkin1953,Rohrlich1953} and radiation~\cite{Bethe1934,Pratt1977} losses. These losses are significant in the case of <<thick>> electrodynamic structures.

A consistent quantitative description of particle-matter interaction can be obtained using Monte-Carlo simulations~\cite{Hebbard1955,Berger1963a,Berger1963b,Berger1964,Perkins1962,Seltzer1974,Seltzer1991,Jenkins1987,Fernandez1993} based on the most rigorous Goudsmit-Sanderson multiple scattering theory~\cite{GS1940}. The algorithms used in numerical calculations should be, on the one hand, sufficiently accurate, and, on the other hand, fast.

In this paper, we describe an approach to modeling the interaction of electron beams with electrodynamic structures in high-current electronic devices. The approach will be validated by calculating electron transmission and reflection coefficients and comparing the obtained results with the numerical simulations~\cite{Berger1963b,Seltzer1974} and experimental data~\cite{Tabata1971,Ito1993}~published in the literature.

After validation, the numerical method will be integrated into a one-dimensional program for modeling high-current devices with an oscillating virtual cathode (VC). We will demonstrate that the scattering of relativistic particles by the anode mesh leads to the formation of an electron cloud near the anode. The cloud particles possess large energy spread and cause significant decrease in the amplitude of field oscillations in vircators. It will be shown that the use of an anode mesh with the thickness approximately equal to the electron mean free path in the anode material leads to decrease in the number of cloud particles and could contribute to increase of the oscillation amplitude.

\section{Passage of relativistic electrons through matter}
\subsection{Monte-Carlo simulation}
The basis of numerical simulation of electron-matter interaction is the Monte-Carlo method ~\cite{Berger1963a,Seltzer1974,Seltzer1991,Jenkins1987}, which can be described as follows. The trajectory of each electron in matter is divided into many small segments. For each segment, the electron energy  is assumed to be constant. When passing from one segment to another, the particle  changes its energy in accordance with the theory of ionization and radiation losses~\cite{Rohrlich1953}. Angular distribution of scattered particle is described by the Goudsmit-Sanderson distribution~\cite{Berger1963a}.
The calculation is carried out until the particle either leaves the substance or loses a significant part of initial energy (in practice, calculations stop when the electron energy  approaches 10 keV).

A significant drawback of the Monte-Carlo method is the computational complexity. To calculate the Goudsmit-Sanderson distribution, it is necessary to sum a large number of terms. Each term contains an integral of a rapidly oscillating function. This circumstance significantly complicates the use of the standard Monte-Carlo method in high-current electronics due to the huge number of particles used in the simulation of high-current devices.

A way out in this situation could be a method, in which the distribution function of the scattering angles would be found using a simple formula. And such a method was proposed in~\cite{Fernandez1993}. It is based on a rigorous theoretical results obtained by Lewis~\cite{Lewis1950}. According to~\cite{Lewis1950}, the most significant mean values depend only on the transport cross-sections: the first one
\begin{equation}
	\sigma_1=2\pi\int_0^\pi(1-\cos\chi)\frac{d\sigma(\chi)}{d\Omega}\sin\chi d\chi
\end{equation}
and second that
\begin{equation}
	\sigma_2=2\pi\int_0^\pi\frac{3}{2}(1-\cos^2\chi)\frac{d\sigma(\chi)}{d\Omega}\sin\chi d\chi.
\end{equation}
Here, $\frac{d\sigma(\chi)}{d\Omega}$ is the differential elastic electron-atom cross-section. For example, the average longitudinal velocity $v$ and the root-mean-square deviation of the transverse velocity in a segment of length $s$ \footnote{The length of $s$ must be much less than the interval over which the particle energy changes significantly.} change in accordance with the formulas as follows, respectively: 
\begin{equation}
	\label{v1}
	v<\cos\theta>=v\exp(-n\sigma_1s)
\end{equation}
and
\begin{equation}
	\label{v2}
	v^2(1-<\cos^2\theta>)=v^2\Big(1-\frac{1+2\exp(-n\sigma_2s)}{3}\Big).
\end{equation}
Symbol $n$ denotes the particle density, $\theta$ is the polar angle between two vectors of particle velocity. The first vector corresponds to the entering to the segment $s$ and the second one corresponds to the segment exit.

The idea stated in~\cite{Fernandez1993} is as follows. If a simple distribution leads to the formulas \eqref{v1} and \eqref{v2}, then the computational complexity of the Monte-Carlo method is significantly reduced. At the same time, the results of modeling particle passage through matter does not change. This was convincingly demonstrated by authors~\cite{Fernandez1993} by calculating the angular distribution of electrons after passed through gold foils of various thicknesses.


\subsection{Reflection and transmission coefficients}
\begin{figure}[ht]
	\begin{center}
		\resizebox{80mm}{!}{\includegraphics{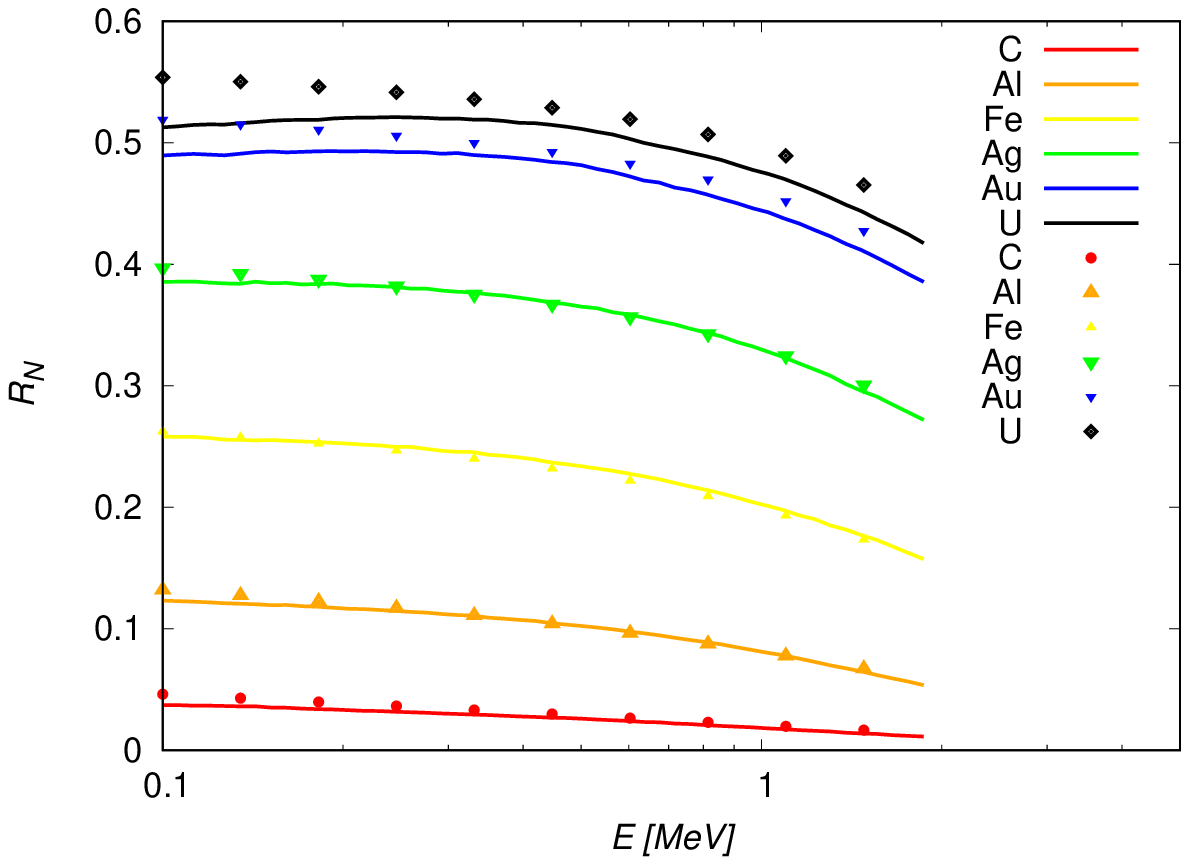}}\resizebox{80mm}{!}{\includegraphics{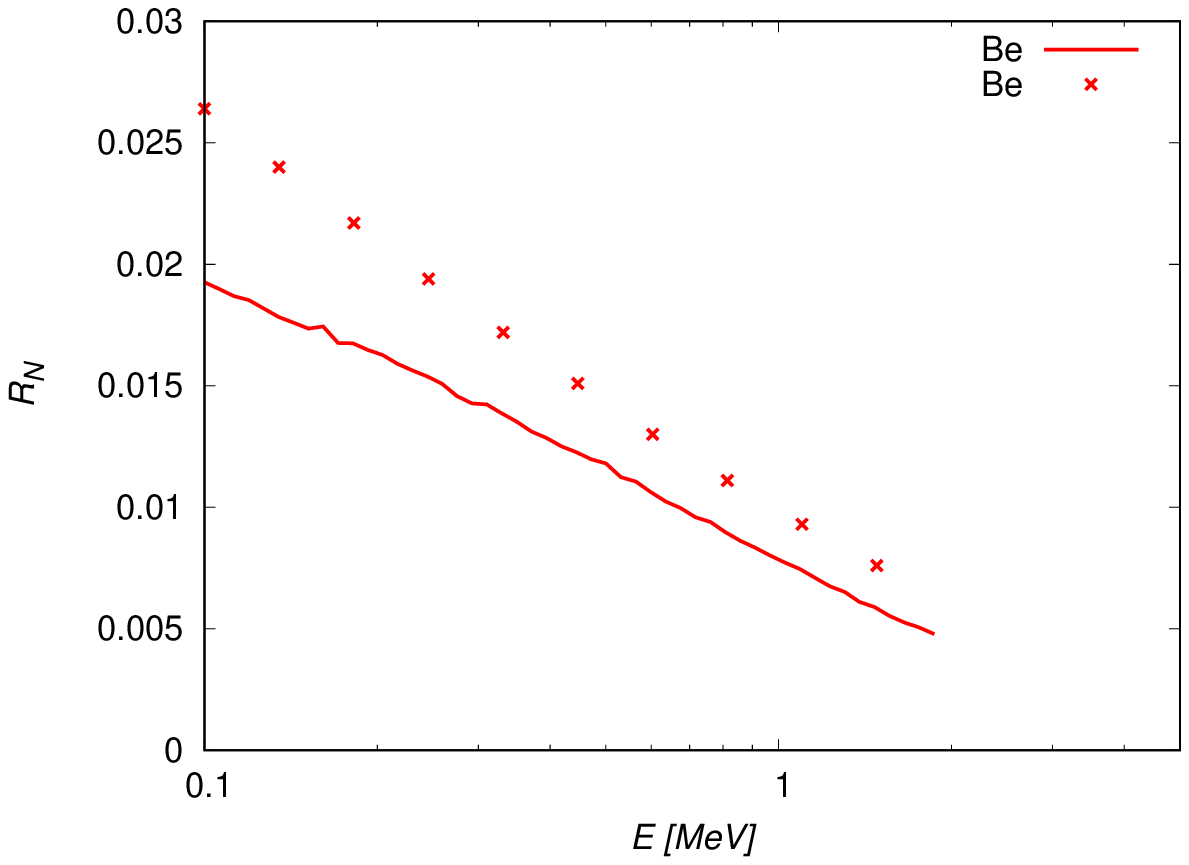}}\\
	\end{center}
	\caption{Electron reflection coefficients: solid lines depict our simulation results, points correspond to  experimental data~\cite{Tabata1971,Ito1993}.} \label{fig:tabata}
\end{figure}

To verify the described approach to simulation of the interaction of electrons with matter, we investigated electron scattering by thick foils made of various materials 
(Be, C, Al, Fe, Ag, Au, U) 
and calculated reflection coefficients.  The energy of particles normally incident on a target varied from 0.1 to 2 MeV. Comparison of our simulation results and experimental data~\cite{Tabata1971,Ito1993} demonstrates better agreement for heavy elements (left plot in figure~\ref{fig:tabata}). 

Simulation results obtained  for light elements 
(right plot in figure~\ref{fig:tabata}) deviate from experimental data in a greater extent: this is due to ignoring the electron-electron collisions in the simulation that increases the angle of multiple scattering by approximately $1+1/Z$ times. The nuclear charge in the denominator indicates that this phenomenon is more significant for light elements.
Difference observed in simulation results and experiment for heavy elements (see curves for gold and uranium in left plot in figure~\ref{fig:tabata}) could be explained by rearrangement of the shells of valence electrons in a solid matter.

\begin{figure}[H]
	\begin{center}
		\resizebox{80mm}{!}{\includegraphics{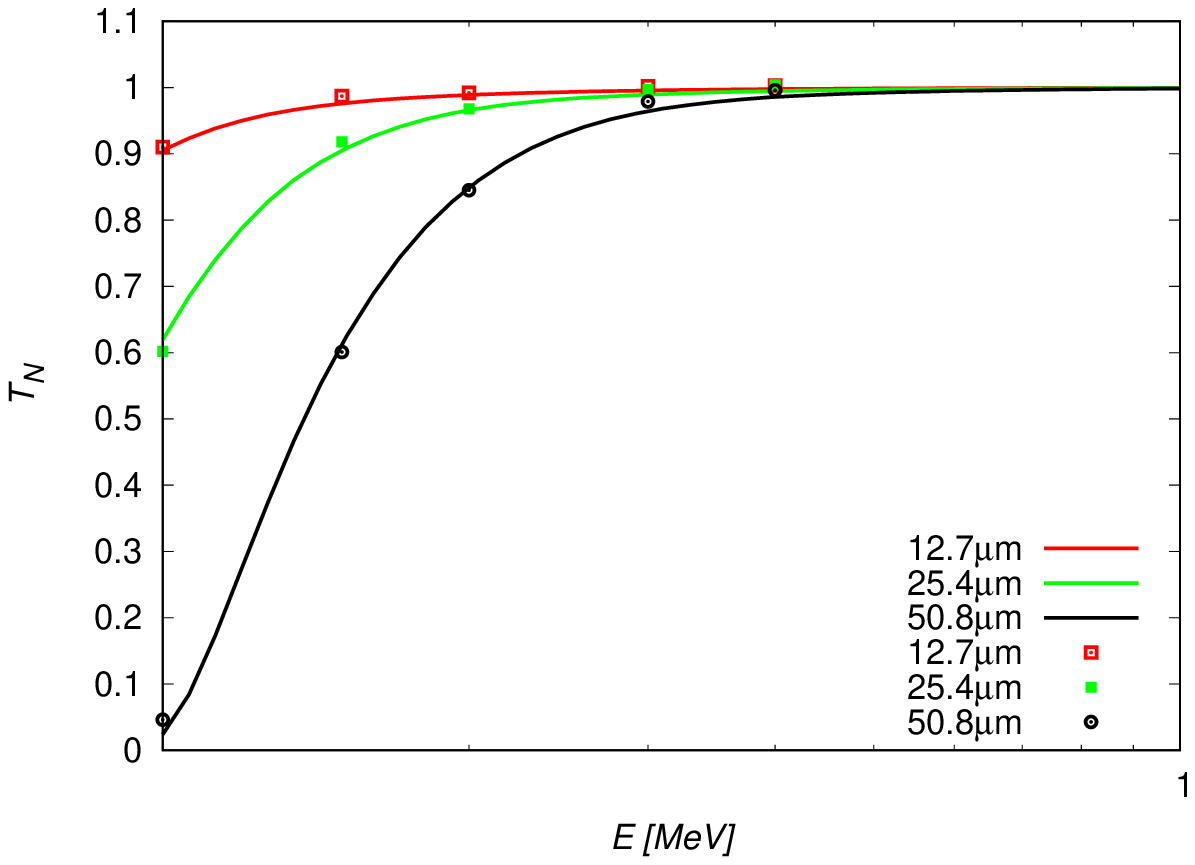}}\resizebox{80mm}{!}{\includegraphics{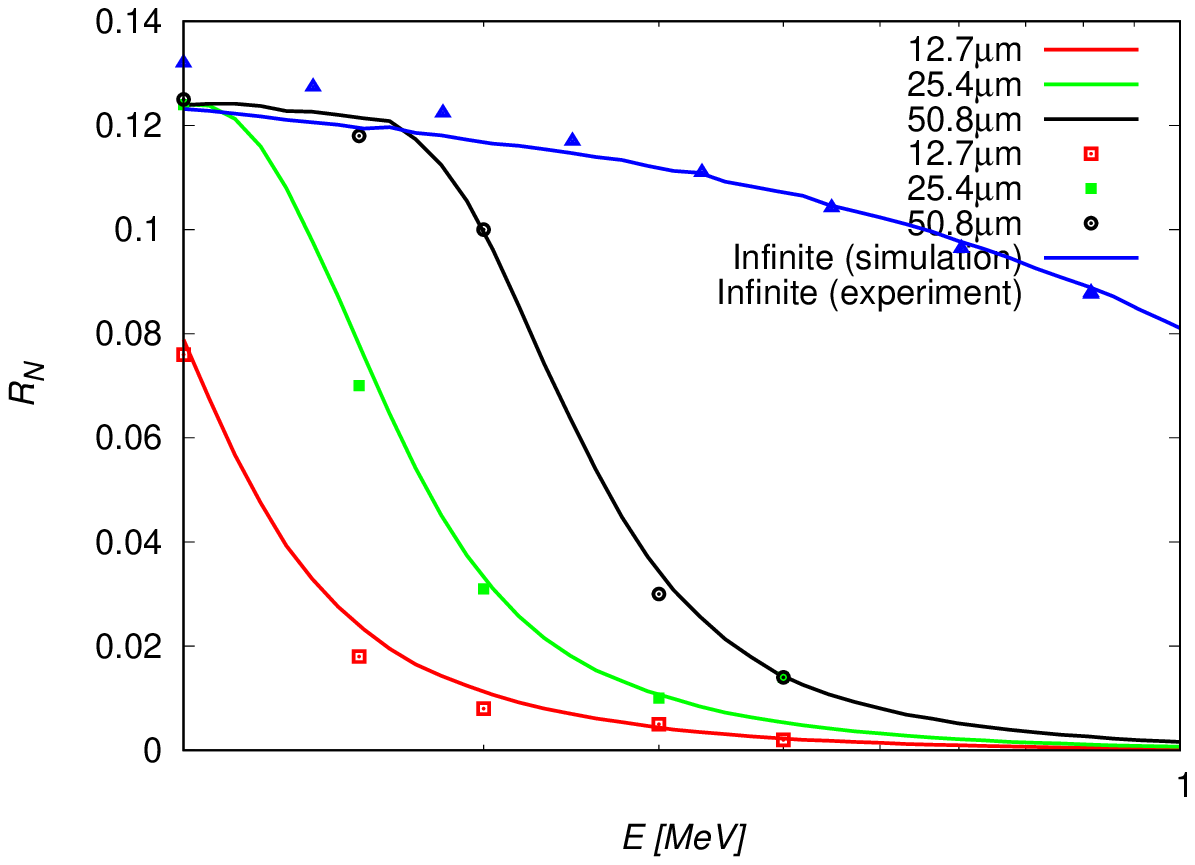}}\\
	\end{center}
	\caption{Transmission (left) and reflection (right) coefficients for electrons scattered by aluminum foils of various thicknesses: solid lines correspond to the simulation results; red, green and black dots are the calculations from the paper\cite{Seltzer1974}; blue dots depict experimental data~\cite{Ito1993}} \label{fig:seltzerAl}
\end{figure}

\begin{figure}[H]
	\begin{center}
		\resizebox{80mm}{!}{\includegraphics{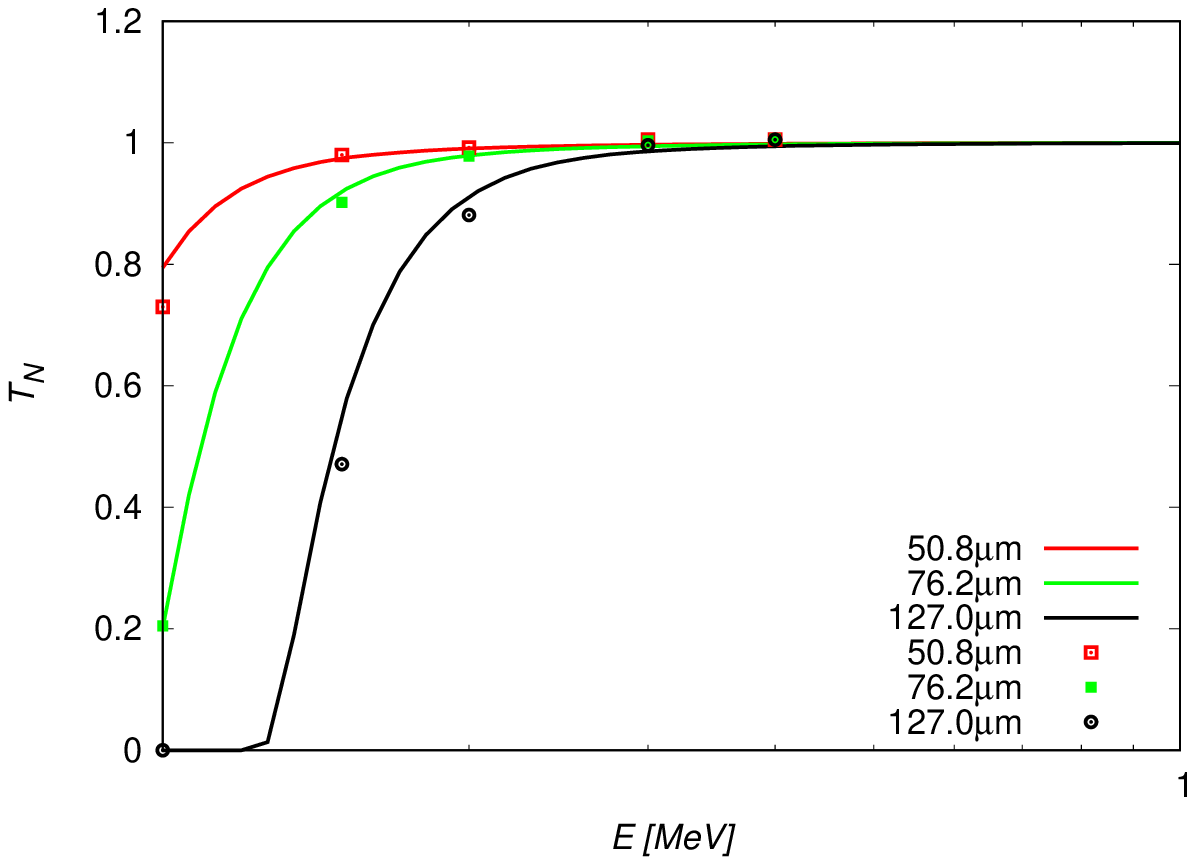}}\resizebox{80mm}{!}{\includegraphics{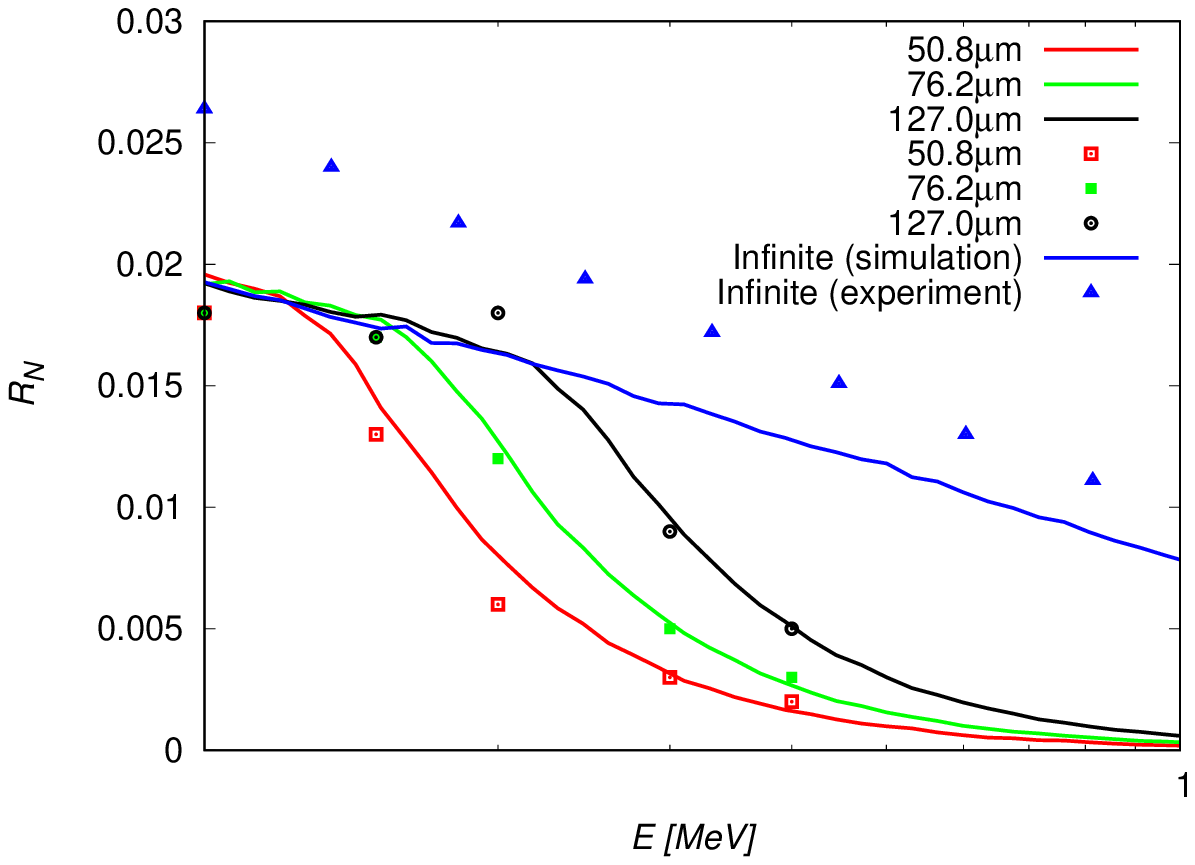}}\\
	\end{center}
	\caption{Transmission (left) and reflection (right) coefficients for electrons scattered by beryllium foils of various thicknesses: solid lines correspond to the simulation results; red, green and black dots are the calculations from the paper\cite{Seltzer1974}; blue dots depict experimental data~\cite{Ito1993}} \label{fig:seltzerBe}
\end{figure}

Figures \ref{fig:seltzerAl}, \ref{fig:seltzerBe} illustrate comparison of simulation results with calculations~\cite{Seltzer1974} for transmission and reflection coefficients for electrons scattered by aluminum and beryllium foils of different thicknesses; comparison with experimental data~\cite{Ito1993} is also provided. Our results are in good agreement with the calculations~\cite{Seltzer1974}.
However, experimentally obtained data for reflection coefficient (right plots in Figures \ref{fig:seltzerAl}, \ref{fig:seltzerBe}) differ from both our simulations and calculations~\cite{Seltzer1974}  that is most likely due to the neglect of electron-electron collisions.

\begin{figure}[H]
	\begin{center}
		\resizebox{80mm}{!}{\includegraphics{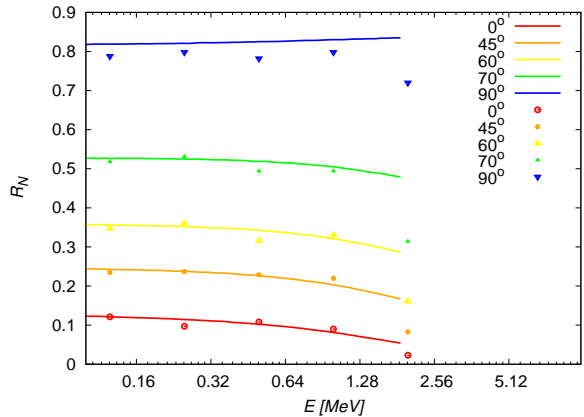}}\\
	\end{center}
	\caption{Dependencies of the electron reflection coefficient on the kinetic energy at different angles of incidence on an aluminum plate: solid lines depict our simulation results and dots show calculation results presented in~\cite{Berger1963b}.} \label{fig:bergerAl}
\end{figure}

Figure \ref{fig:bergerAl} shows the dependence of the reflection coefficient on the kinetic energy of a particle incident on an aluminum plate. Three solid curves present the simulation results obtained for different angles of incidence. At energy values up to $\sim2$~MeV our simulation results are  in good agreement with both calculation~\cite{Berger1963b} and experimental data ~\cite{Ito1993}. At the energy $\sim2$~MeV, the reflection coefficients given in~\cite{Berger1963b} demonstrate a noticeable decrease in contrast to our calculations. 
Such behavior applies to the case of normal incidence. 

Slight deviation of our simulation results from those presented in Berger's paper~\cite{Berger1963b} could be associated with different approaches to
the consideration of multiple scattering. We used the approach \cite{Fernandez1993} based on rigorous results obtained by Lewis \cite{Lewis1950} from the Goudsmit-Sanderson theory, while Berger resorted to the approximate Moli$\grave{e}$re theory. 
Note that experimental data obtained at normal incidence ($\theta=0^{o}$) (figure~\ref{fig:tabata}) does not show such a decrease in the reflection coefficient.

Thus, the simulation results obtained in this work demonstrate good agreement with experimental data and numerical calculations published in the literature. Some discrepancy ($\sim20\%$) with  the experimental data is observed in the reflection coefficient for electrons with energy $\sim0.1$~MeV scattered by beryllium foil.

\section{VC oscillations}
\begin{figure}[ht]
	\begin{center}
		\resizebox{80mm}{!}{\includegraphics{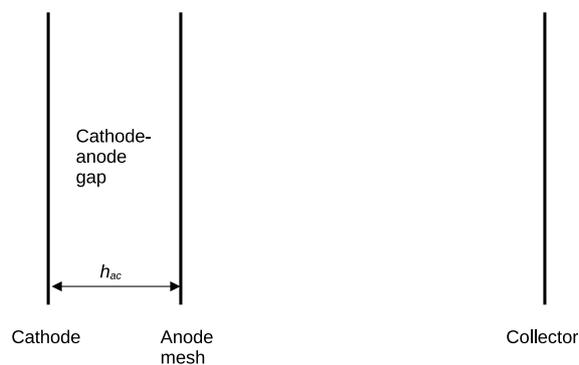}}\\
	\end{center}
	\caption{One-dimensional model of a high-current device with an oscillating VC.} 
	\label{fig:vircator}
\end{figure}

The approach developed to simulate the passage of electrons through matter was integrated into a one-dimensional code designed to simulate high-current devices with an oscillating virtual cathode. 
This code calculates motion of relativistic electrons in a self-consistent longitudinal electric field by the particle-in-cell method. The computational domain consists of two parts (figure \ref{fig:vircator}): the cathode-anode gap and the drift space. The potential of the right wall of the drift space can be either equal to the anode potential (in the case of a vircator) or to the cathode potential (in the case of a reflex triode). Injection of particles into the system is carried out under conditions of unlimited emission capability of the cathode. This condition corresponds to the regime of explosive electron emission that takes place in high-current accelerators. 

\begin{figure}[ht]
	\begin{center}
		\resizebox{80mm}{!}{\includegraphics{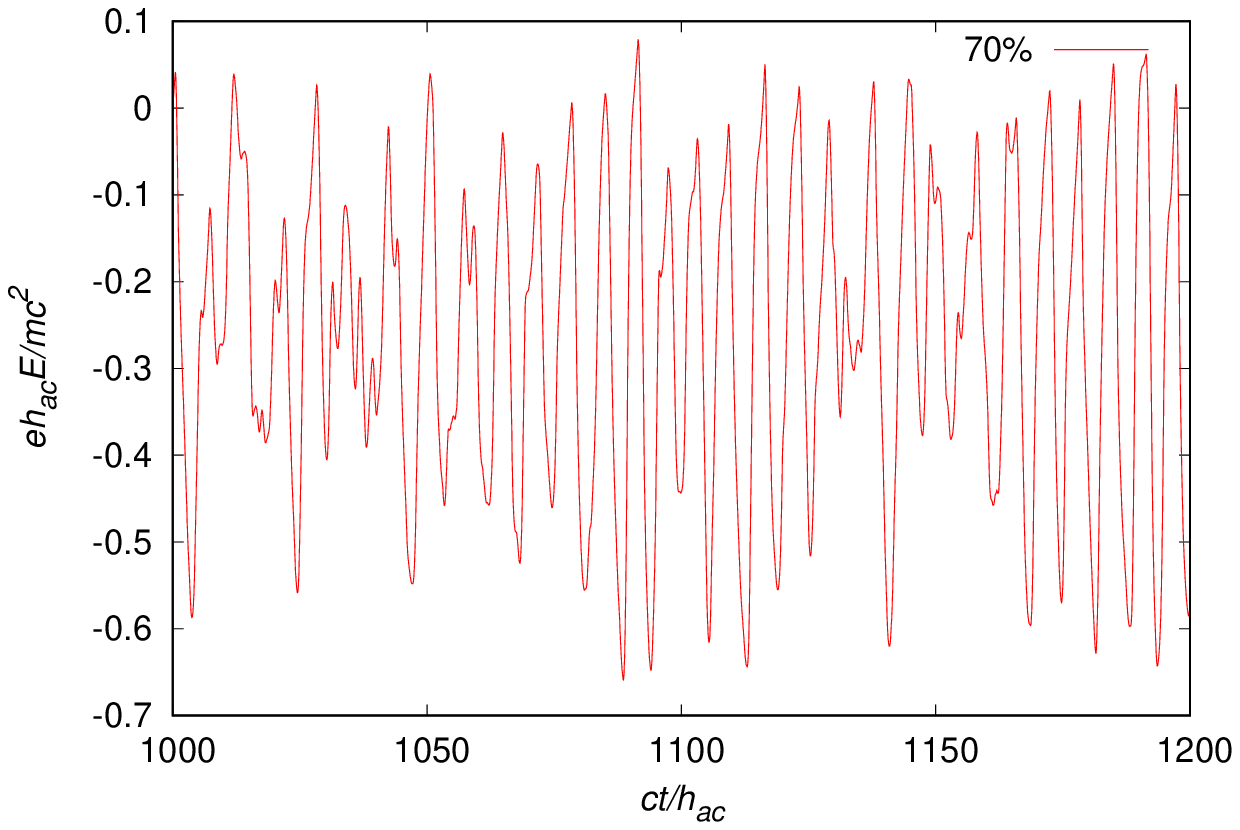}}\resizebox{80mm}{!}{\includegraphics{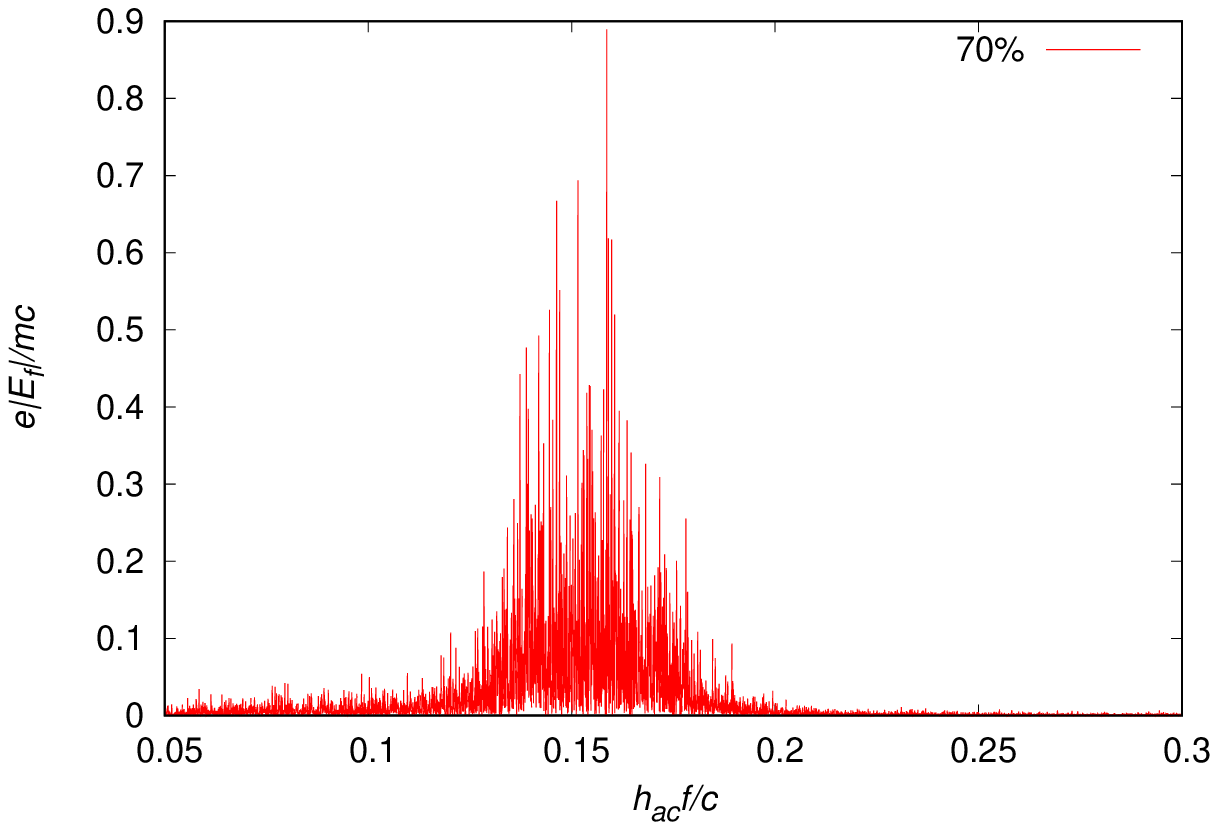}}\\
	\end{center}
	\caption{Electric field in the VC region and electric field spectrum. Absorption model.} \label{fig:absorbtion}
\end{figure}

\begin{figure}[ht]
	\begin{center}
		\resizebox{80mm}{!}{\includegraphics{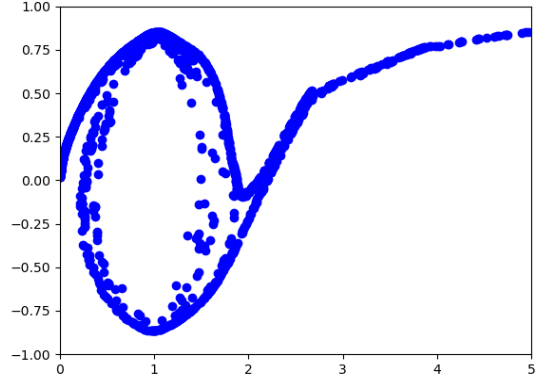}}\\
	\end{center}
	\caption{Phase portrait of an electron beam 
		in the absorption model. The longitudinal coordinates and particle velocities are shown along the abscissa in the units of the cathode-anode gap and along the ordinate in units of speed of light, respectively.} \label{fig:absorbtionmovie}
\end{figure}

The initial version of the PIC code took no account of the scattering by anode mesh: the anode was assumed to be semitransparent and was characterized by a single parameter, the geometric transparency $\eta$ which varied from zero to one. When a particle passed through the anode, its charge was multiplied by the transparency coefficient. The procedure corresponded to the partial absorption of particles by the anode mesh. Neither the reflection of particles, nor the deceleration of particles in the anode material, nor  spread of particle velocity  due to multiple scattering were taken into account by the code. In what follows, we will refer to the simulation model just described as the absorption model.
Figures \ref{fig:absorbtion} and \ref{fig:absorbtionmovie} show the electric field oscillations, spectrum, and the phase portrait of the beam in the vircator in the absorption model. The geometric transparency is $\eta=0.7$.

In the updated version of PIC code the scattering of particles  by the anode mesh was added via consideration of electron-matter interaction. (We will refer to the new simulation model as the scattering model.) The anode mesh was described by three parameters: the geometric transparency of the mesh, the anode material, and anode thickness $d$. The geometric transparency $\eta$ was introduced in terms of probability ($1-\eta$) for a particle to hit the anode mesh, which was supposed to be a steel foil of thickness $d$ with holes. The ratio of the sum of hole areas to the total foil area defined the geometric transparency $\eta$.

Thus, a particle passing through the anode is scattered with probability $1-\eta$ rather than $\eta$. Particle scattering by the anode is calculated exactly in the same way as scattering by a solid foil\footnote{The possible complex structure of the anode is not taken into account. For example, the mesh may be made of round wires, which cannot be described in terms of a one-dimensional model.}. The scattered (unabsorbed) particle enters either the cathode-anode gap or the drift space with a random component of the transverse momentum and with a reduced longitudinal velocity due to ionization energy loss. In this case, time of particle motion inside the anode material is neglected. Since the characteristic thickness $d$ is much smaller than the characteristic value of the cathode-anode gap, we neglect time, which particle spends  inside the metal. 

\begin{figure}[H]
	\begin{center}
		\resizebox{80mm}{!}{\includegraphics{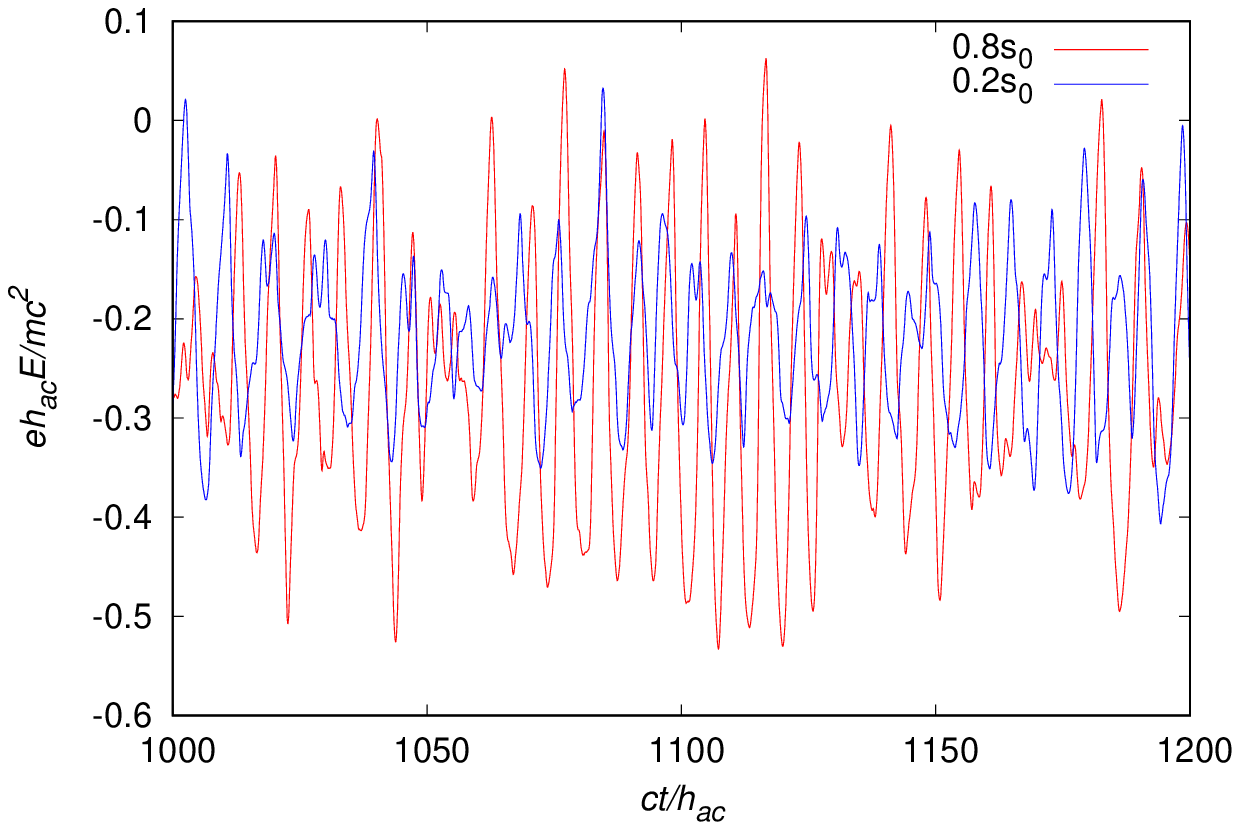}}\resizebox{80mm}{!}{\includegraphics{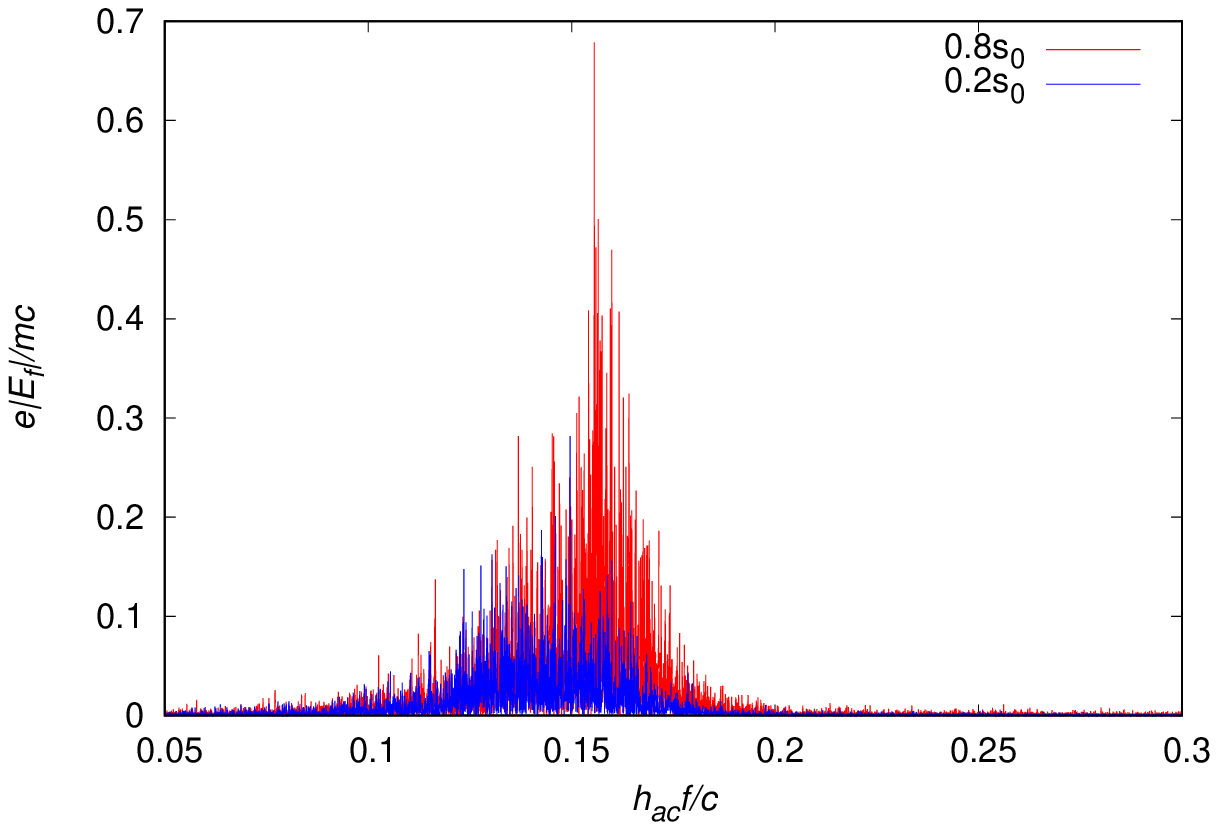}}\\
	\end{center}
	\caption{Electric field and its spectrum in the VC region.} \label{fig:iron}
\end{figure}

\begin{figure}[H]
	\begin{center}
		\resizebox{80mm}{!}{\includegraphics{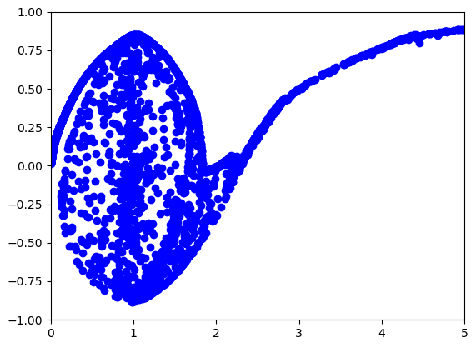}}\resizebox{80mm}{!}{\includegraphics{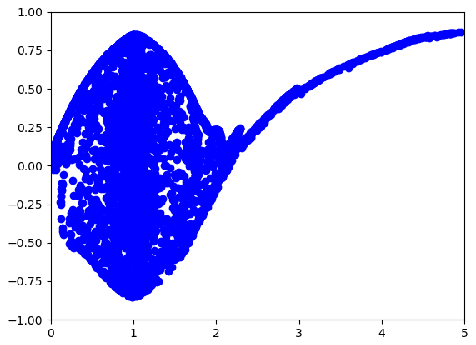}}\\
	\end{center}
	\caption{Phase portraits of the electron beam at different thicknesses of the steel anode. The phase portraits shown on the left and on the right correspond to the thickness $0.8s_0$ and $0.2s_0$, respectively. The longitudinal coordinates and particle velocities are shown along the abscissa in the units of the cathode-anode gap and along the ordinate in units of speed of light, respectively} \label{fig:ironmovie}
\end{figure}

Figures \ref{fig:iron} and \ref{fig:ironmovie} show the electric field oscillations, spectra and the phase portraits in the case of steel anode of different thicknesses (thicknesses are indicated on the plot in units of the mean free path $s_0$). The geometric transparency of anode mesh is $\eta=0.7$.
Simulation of vircators showed that  electron scattering by the anode mesh leads to the formation of a cloud of low-energy electrons with a large spread in velocities. These electrons appear due to loss of longitudinal momentum after scattering by the anode mesh. Due to deceleration, these electrons do not have enough energy to reach the either real or virtual cathode. As a result, they oscillate in the potential well between the cathode and the virtual cathode until complete absorption. Scattering by the anode leads to a random change in phase of electron oscillations. This change prevents particles from generating coherent oscillations. In addition, an electron cloud partially shields the anode and blocks the vacuum diode. As a consequence, the amount of electrons, which could participate in collective oscillations, appears smaller. As a result, the amplitude of oscillations in a vircator decreases.

The vircator operation can be improved if the thickness of the partitions in the anode mesh is chosen to be close to the electron path length in the anode material. In this case, the number of scattered electrons passing through the anode could be significantly reduced. As a consequence, fewer particles in the cloud would affect oscillation amplitude decreasing.

\section{Сonclusion}
Multiple scattering of a relativistic electron beam by an electrodynamic structure 
accompanied by ionization and radiation energy losses significantly influences on operation of a high-current virtual cathode oscillator. 
In a  vircator, the scattering of a high-current relativistic electron beam by the anode mesh leads to formation of an electron cloud near the anode. The cloud particles possess low energy and large spread in velocities caused by multiple scattering and ionization losses. The electrons captured by the cloud do not participate in the oscillations of the virtual cathode and partially block a vircator. As a result, the amplitude of the electric field oscillations is reduced. In order to increase the oscillation amplitude, the thickness of wires forming the anode mesh should be equal to the mean free path for electrons in the mesh material. Use of anode mesh corresponding the above requirements enables to minimize the number of scattered electrons passing through the anode.

We applied the approach for modeling of multiple scattering described by~\cite{Fernandez1993} in the simulation code developed by us to calculate the influence of electron ionization losses  and multiple scattering by the anode mesh on particles' dynamics in a vircator.  The above approach significantly reduces calculation time as compared with the most popular concept based on the Goudsmit-Sanderson distribution.

\label{last}
\end{document}